
\input lanlmac.tex
\input epsf
\font\cmss=cmss10 \font\cmsss=cmss10 at 7pt
\ifx\epsfbox\UnDeFiNeD\message{(NO epsf.tex, FIGURES WILL BE IGNORED)}
\def\figin#1{\vskip2in}
\else\message{(FIGURES WILL BE INCLUDED)}\def\figin#1{#1}\fi
\def\tfig#1{{\xdef#1{Fig.\thinspace\the\figno}}
Fig.\thinspace\the\figno \global\advance\figno by1}
%


\def\ie{{\it i.e.}}
\def\eg{{\it e.g.}}
\def\subsubsection#1{

\vskip 0.3cm
{\it #1}
\vskip 0.2cm

}

\def\inbar{\,\vrule height1.5ex width.4pt depth0pt}
\def\ie{{\it i.e.}}

\def\Tr{\,{\rm Tr}\, }

\def\Re{\,{\rm Re}\, }

\def\rangl{\right\rangle   }
\def\langl{\left\langle  }
\def\({\left(}
\def\){\right)}
\def\[{\left[}
\def\]{\right]}
\def\p{\partial}

\def\11{1\!\! 1}

\def\hf{{1\over 2}}
\def\IR{\relax{\rm I\kern-.18em R}}
\def\IC{{\relax\hbox{$\inbar\kern-.3em{\rm C}$}}}


\def\eps{\varepsilon}

\def\l{\lambda}

\def\s{\sigma}

\def\G{\Gamma}

\def\U{\Upsilon}

 \def\IZ{\relax\ifmmode\mathchoice
{\hbox{\cmss Z\kern-.4em Z}}{\hbox{\cmss Z\kern-.4em Z}}
{\lower.9pt\hbox{\cmsss Z\kern-.4em Z}}
{\lower1.2pt\hbox{\cmsss Z\kern-.4em Z}}\else{\cmss Z\kern-.4em
Z}\fi}
\def\CA {{\cal A}}

\def\CF {{\cal F}}

\def\CL {{\cal L}}

\def\CT {{\cal T}}


\def\XX{X}
\def\mul{\mu_{_L}}

\def\efo{e^{-{1\over R}\XX}}
\def\ego{e^{-{R-1\over R}\XX}}

\def\under#1#2{\mathop{#1}\limits_{#2}}

\def\FSL{\CF}

 \def\R{\relax{\rm I\kern-.18em R}}
\font\cmss=cmss10 \font\cmsss=cmss10 at 7pt
\def\Z{\relax\ifmmode\mathchoice
{\hbox{\cmss Z\kern-.4em Z}}{\hbox{\cmss Z\kern-.4em Z}}
{\lower.9pt\hbox{\cmsss Z\kern-.4em Z}}
{\lower1.2pt\hbox{\cmsss Z\kern-.4em Z}}\else{\cmss Z\kern-.4em Z}\fi}

\def\np#1#2#3{{\it Nucl. Phys.} {\bf B#1} (#2) #3}
\def\pl#1#2#3{{\it Phys. Lett.} {\bf B#1} (#2) #3}
\def\prl#1#2#3{{\it Phys. Rev. Lett.} {\bf #1} (#2) #3}
\def\physrev#1#2#3{{\it Phys. Rev.} {\bf D#1} (#2) #3}

\def\mpl#1#2#3{{\it Mod. Phys. Lett.} {\bf #1} (#2) #3}
\def\ijmp#1#2#3{{\it Int. J. Mod. Phys.} {\bf #1} (#2) #3}

\def\jhep#1#2#3{{\it JHEP} {\bf #1} (#2) #3}

%
\lref\PolchinskiFQ{
J.~Polchinski,
``Combinatorics Of Boundaries In String Theory,''
\physrev{50}{1994}{6041}, hep-th/9407031.
}
\lref\FukumaTJ{
M.~Fukuma and S.~Yahikozawa,
``Comments on D-instantons in $c < 1$ strings,''
Phys.\ Lett.\ B {\bf 460}, 71 (1999)
[arXiv:hep-th/9902169].
}

\lref\NevesXT{
R.~Neves,
``D-instantons in non-critical open string theory,''
\pl{411}{1997}{73},
hep-th/9706069.
}

\lref\MARTINEC{
E. J. Martinec, ``The Annular Report on Non-Critical String Theory'',
hep-th/0305148.}
\lref\ZamolodchikovAH{
A.~B.~Zamolodchikov and A.~B.~Zamolodchikov,
``Liouville field theory on a pseudosphere,''
hep-th/0101152.
}
\lref\EynardSG{
B.~Eynard and J.~Zinn-Justin,
``Large order behavior of 2-D gravity coupled to d < 1 matter,''
\pl{302}{1993}{396}, hep-th/9301004.
}

\lref\DijkgraafHK{
R.~Dijkgraaf, G.~W.~Moore and R.~Plesser,
``The Partition function of 2-D string theory,''
\np{394}{1993}{356}, 
hep-th/9208031.
}

\lref\GinspargCY{
P.~Ginsparg and J.~Zinn-Justin,
``Action Principle And Large Order Behavior Of Nonperturbative Gravity,''
{\it Proc. of Cargese Workshop: 
Random Surfaces and Quantum Gravity, Ed. by O. Alvarez, et al., Cargese, 
France, May 27 - June 2, 1990}
}

\lref\KazakovPM{
V.~Kazakov, I.~K.~Kostov and D.~Kutasov,
``A matrix model for the two-dimensional black hole,''
\np{622}{2002}{141},
hep-th/0101011.
}
\lref\CARDY{   J.L. Cardy, ``Boundary conditions, fusion rules and the
Verlinde formula'', \np{324}{1989}{581}.}
\lref\ZamolodchikovAA{
A.~B.~Zamolodchikov and A.~B.~Zamolodchikov,
``Structure constants and conformal bootstrap in Liouville field theory,''
\np{477}{1996}{577},
hep-th/9506136.
}
\lref\DornXN{
H.~Dorn and H.~J.~Otto,
``Two and three point functions in Liouville theory,''
\np{429}{1994}{375},
hep-th/9403141.
}
\lref\MooreGA{
G.~Moore,
``Gravitational phase transitions and the Sine-Gordon model,''
hep-th/9203061.
}
\lref\HsuCM{
E.~Hsu and D.~Kutasov,
``The Gravitational Sine-Gordon model,''
\np{396}{1993}{693}, hep-th/9212023.
}

\lref\DouglasVE{
M.~R.~Douglas and S.~H.~Shenker,
``Strings In Less Than One-Dimension,''
\np{335}{1990}{635}.
}
\lref\BrezinRB{
E.~Brezin and V.~A.~Kazakov,
``Exactly Solvable Field Theories Of Closed Strings,''
\pl{236}{1990}{144}.
}
\lref\GrossVS{
D.~J.~Gross and A.~A.~Migdal,
``Nonperturbative Two-Dimensional Quantum Gravity,''
\prl{64}{1990}{127}.
}

\lref\McGreevyKB{
J.~McGreevy and H.~Verlinde,
``Strings from tachyons: The c = 1 matrix reloaded,''
hep-th/0304224.
}

\lref\KlebanovKM{
I.~R.~Klebanov, J.~Maldacena and N.~Seiberg,
``D-brane decay in two-dimensional string theory,''
hep-th/0305159.
}

\lref\DavidSK{
F.~David, ``Phases Of The Large N Matrix Model And
Nonperturbative Effects In 2-D Gravity,''
\np{348}{1991}{507}.
}

\lref\DavidZA{
F.~David,
``Nonperturbative effects in matrix models and vacua
of two-dimensional gravity,''
\pl{302}{1991}{403}, hep-th/9212106.
}
\lref\BoulatovXZ{
D.~Boulatov and V.~Kazakov, 
``One-dimensional string theory with vortices 
as the upside down matrix oscillator'',  preprint LPTENS 91/24 (1991),
Int.\ J.\ Mod.\ Phys.\ A {\bf 8}, 809 (1993)
hep-th/0012228.}
\lref\HoppeXG{
J.~Hoppe, V.~Kazakov and I.~K.~Kostov,
``Dimensionally reduced SYM(4) as solvable matrix quantum mechanics,''
Nucl.\ Phys.\ B {\bf 571}, 479 (2000)
hep-th/9907058.
}

\lref\RABI{S. Elitzur, A. Forge and E. Rabinovici,
\np {\bf B359} (1991) 581. }
\lref\WADIA{G. Mandal, A. Sengupta, and  S. Wadia,
Mod. Phys. Lett. {\bf A6} (1991) 1685.}
\lref\WITTEN{E. Witten, Phys. Rev. {\bf D44} (1991) 314.}
\lref\FZ{Al. B. Zamolodchikov, unpublished;
V.A. Fateev, \pl {\bf B357} (1995) 397.}
\lref\FZZ{V. Fateev, A. Zamolodchikov and  Al. Zamolodchikov,
unpublished.}
\lref\DDK{V. Knizhnik, A. Polyakov and A. Zamolodchikov,
\mpl{A3}{1988}{819};
F. David, \mpl{A3}{1988}{1651};
J. Distler and H. Kawai, \np{321}{1989}{509}.}
\lref\KAZMIG{V. Kazakov and A. A. Migdal, \np{311}{1988}{171}.}
\lref\GRKL{D. Gross and I. Klebanov, \np{344}{1990}{475}.}
\lref\GRKLbis{D. Gross and I. Klebanov, \np{354}{1990}{459}.}

\lref\BULKA{D. Boulatov and V.Kazakov, 
 preprint LPTENS 91/24 (1991),
\ijmp{A8}{1993}{809}, revised version: hep-th/0012228.}
\lref\DASJP{S. Das and A. Jevicki,
\mpl{A5}{1990}{1639}.}
\lref\DASWD{
S. R. Das, Mod. Phys. Lett. {\bf A8} (1993) 69, 1331;
A. Dhar, G. Mandal and S. Wadia, Mod. Phys. Lett. {\bf A7} (1992) 3703;
{\bf A8} (1993) 1701; A. Dhar \np {\bf B507} (1997) 277.}
\lref\JEV{ A. Jevicki and D. Yoneya, Nucl. Phys. {\bf B411} (1994) 64.}
\lref\MATYTSIN{A. Matytsin and P. Zaugg, hep-th/9611170,
\np {\bf B497} (1997) 658; hep-th/9701148, \np {\bf B497} (1997) 699.}
\lref\MOORE{ G. Moore, hep-th/9203061.}
\lref\On{I. Kostov and M. Staudacher,  \np {\bf B384} (1992) 459.}
\lref\JM{M. Jimbo and T. Miwa, \ Publ. RIMS, Kyoto Univ. {\bf 19}, No. 3
(1983) 943.}
\lref\Hir{R. Hirota,  Direct Method in Soliton Theory, {\it Solitons},
Ed. by R. K. Bullogh and R. J. Caudrey, Springer, 1980.}
\lref\HK{E. Hsu and  D. Kutasov,  hep-th/9212023, Nucl. Phys.
{\bf B396} (1993) 693.}
\lref\UT{K. Ueno and K. Takasaki, Adv. Stud. Pure Math. {\bf 4} (1984) 1;
K. Takasaki, Adv. Stud. Pure Math. {\bf 4} (1984) 139.}
\lref\SAKA{S. Kakei, ``Toda lattice hierarhy and Zamolodchikov's
conjecture", solv-int/9510006 }
\lref\PBM{A. Prudnikov, Yu. Brichkov, O. Marychev, ``Integrals and
Series'', Nauka , Moscow, 1981.}
\lref\DVV{R. Dijkgraaf, E. Verlinde and H. Verlinde,
Nucl. Phys. {\bf B371} (1992) 269.}
\lref\HS{G. Horowitz and A. Strominger, Nucl. Phys. {\bf B360}
(1991) 197; J. Maldacena and A. Strominger, hep-th/9710014.}
\lref\OV{H. Ooguri and C. Vafa, hep-th/9511164, Nucl. Phys.
{\bf B463} (1996) 55.}
\lref\GiveonPX{
A.~Giveon and D.~Kutasov,
``Little string theory in a double scaling limit,''
\jhep{9910}{1999}{034},
hep-th/9909110.
}
\lref\GiveonTQ{
A.~Giveon and D.~Kutasov,
``Comments on double scaled little string theory,''
\jhep{0001}{2000}{023},
hep-th/9911039.
}
\lref\TESCH{J. Teschner, hep-th/9712256, Nucl. Phys. {\bf B546}
(1999) 390; hep-th/9712258, Nucl. Phys. {\bf B546} (1999) 369;
hep-th/9906215, Nucl. Phys. {\bf B571} (2000) 555.}
\lref\difkut{P. Di Francesco and D. Kutasov, hep-th/9109005,
Nucl. Phys. {\bf B375} (1992) 119.}
\lref\FATA{ L. D. Fadeev and L. A. Takhtajan, ``Hamiltonian Methods in
the Theory of Solitons'', Springer-Ferlag (1987).}

\lref\ShenkerUF{
S.~H.~Shenker,
``The Strength Of Nonperturbative Effects In String Theory,''
{\it Presented at the Cargese Workshop on Random Surfaces, Quantum Gravity and Strings, Cargese, France, May 28 - Jun 1, 1990}
}
\lref\BRKA{E. Brezin, V. Kazakov and Al. Zamolodchikov,
\np {\bf B338}(1990) 673.}
\lref\PARISI{ G. Parisi, \pl {\bf B238} (1990) 209, 213.}
\lref\GRMI{ D. Gross and N. Miljkovic, \pl {\bf B238} (1990) 217.}
\lref\GIZI{P. Ginsparg and J. Zinn-Justin, \pl{240}{1990}{333}.}
\lref\POLCHINSKI{ J. Polchinski, ``What is String Theory?'',
Lectures presented at the 1994 Les Houches Summer School ``Fluctuating
Geometries in Statistical Mechanics and Field Theory'', hep-th/9411028. }
\lref\GP{G. Gibbons and M. Perry,  hep-th/9204090, Int. J. Mod. Phys.
{\bf D1} (1992) 335; C. R. Nappi and A. Pasquinucci, hep-th/9208002,
Mod. Phys. Lett. {\bf A7} (1992) 3337.}
\lref\YK{I. Kogan, {\it JETP Lett.} 44 (1986) 267, 45 (1987) 709.}
\lref\DOUGLAS{M. R. Douglas,  ``Conformal theory techniques
in Large $N$ Yang-Mills Theory'', talk at the 1993 Carg\`ese meeting,
hep-th/9311130.}
\lref\KLEBANOV{I. Klebanov, proceedings of the ICTP
Spring School on String Theory and Quantum Gravity,
 Trieste, April 1991, hep-th/9108019.}
\lref\KAZREV{V. Kazakov,  ``Bosonic strings and string field theories
in one-dimensional target space'', proceedings of
Cargese workshop on Random Surfaces, Quantum Gravity and Strings,
1990.}
\lref\ginsmoore{P. Ginsparg and G. Moore, proceedings of TASI
1992, hep-th/9304011.}
\lref\trtr{ I. Klebanov, \physrev{51}{1995}{1836}. }
\lref\DMP{R. Dijkgraaf, G. Moore, R. Plesser, \np{394}{1993}{356}.}
\lref\adscft{J. Maldacena, hep-th/9711200, Adv. Theor. Math. Phys.
{\bf 2} (1998) 231.}
\lref\KAZMA{ V. A. Kazakov, Solvable Matrix Models, proceedings of the
MSRI Workshop ``Matrix Models and Painlev\'e Equations'', Berkeley
(USA) 1999; hep-th/0003064.}
\lref\KS{D. Kutasov and N. Seiberg, Nucl. Phys. {\bf B358} (1991) 600.}
\lref\niarchos{V. Niarchos, hep-th/0010154.}
%
%
\rightline{SPHT-t03/079, LPTENS-03/21, EFI-03-29}

\vskip -0.5 cm

\Title{}
{\vbox{\centerline { Non-Perturbative Effects }
\centerline{ in Matrix  Models and  D-branes }
}}
%
%
\vskip -0.5cm

\centerline{Sergei Yu. Alexandrov,$^{12}$\footnote{$^{\ast}$}
{alexand@spht.saclay.cea.fr} \
Vladimir A. Kazakov$^{1}$\footnote{$^{\circ}$}
{{kazakov@physique.ens.fr}}  \
and David Kutasov$^{3}$\footnote{$^{\dagger}$}
{kutasov@theory.uchicago.edu}}
\bigskip
{ \ninepoint
\centerline{$^1${\it  Laboratoire de Physique Th\'eorique de l'Ecole
Normale Sup\'erieure,\footnote{$^\ast$}{Unit\'e mixte de Recherche du
Centre National de la Recherche Scientifique et de  l'Ecole Normale
Sup\'erieure associ\'ee \`a l'Universit\'e de Paris-Sud et 
l'Universit\'e  Paris-VI} }}
\centerline{{\it  24 rue Lhomond, 75231 Paris CEDEX, France}}
\centerline{$^2${\it Service de Physique Th\'eorique,
CNRS - URA 2306, C.E.A. - Saclay,}}
\centerline{  \it F-91191 Gif-Sur-Yvette CEDEX, France}
\centerline{$^3${\it Department of Physics, University of Chicago,}}
\centerline{\it 5640 S. Ellis Avenue, Chicago, IL 60637, USA}
}

 \vskip 1cm

\baselineskip12pt{
\noindent

The large order growth of string perturbation theory in $c\le1$ conformal field
theory coupled to world sheet gravity implies the presence of $O(e^{-{1\over g_s}})$
non-perturbative effects, whose leading behavior can be calculated in the matrix
model approach. Recently it was proposed that the same effects should be
reproduced by studying certain localized D-branes in Liouville Field Theory, 
which were constructed by A. and Al. Zamolodchikov. We discuss this 
correspondence in a number of different cases: unitary minimal models 
coupled to Liouville, where we compare the continuum analysis to the 
matrix model results of Eynard and Zinn-Justin, and
compact $c=1$ CFT coupled to Liouville in the presence of a condensate
of winding modes, where we derive the matrix model prediction and compare
it to Liouville theory. In both cases we find agreement between the
two approaches. The $c=1$ analysis also leads to predictions about
properties of D-branes localized in the vicinity of the tip of the cigar
in $SL(2)/U(1)$ CFT with $c=26$.


\Date{June, 2003}

\baselineskip=14pt plus 2pt minus 2pt


\newsec{Introduction }

In the late 1980's and early 1990's it was pointed out that low dimensional
models of non-critical string theory, corresponding to $c\le 1$ conformal
field theories coupled to world sheet quantum gravity, provide interesting toy
models in which one can hope to study non-perturbative effects in string theory
in a controlled setting. In particular, in \refs{\DouglasVE,\BrezinRB,\GrossVS}
it was shown that certain large $N$ matrix models in a double scaling limit allow
one to efficiently compute amplitudes in these theories to all orders in string
perturbation theory. The general structure of the perturbative amplitudes is
\eqn\ampl{
\CA\simeq \sum_{h=0}^\infty g_s^{2h-2} c_h~,
}
where the genus $h$ contribution $c_h$ can be computed from the matrix model.

The question of the non-perturbative completion of these perturbative results
was raised already in the original papers on the double scaling limit
and was studied further in
\refs{\DavidSK,\DavidZA,\EynardSG} and other papers. The matrix model results
for the coefficients $c_h$ showed that the latter grow with the genus like
$c_h\simeq a^h(2h)!$; $a$ was found to be positive for unitary models,
so that the amplitudes are not Borel summable. Standard results on asymptotic
series suggest that in this situation
the leading non-perturbative ambiguities are of order
$\CA_{\rm np}\sim  g_s^{f_A}\exp(-f_D/g_s)$, where $f_A$
and $f_D$ are computable from the large order behavior of the perturbative series.

In another significant development, S. Shenker  \ShenkerUF\ pointed out that the
$(2h)!$ large order behavior exhibited by low-dimensional non-critical strings is 
in fact expected to be generic in string theory, and thus the $\exp(-1/g_s)$
non-perturbative effects are very general as well. A natural source of such effects
was found by J. Polchinski \PolchinskiFQ, who suggested that when going beyond perturbation
theory in closed string theory, one has to include contributions of Riemann
surfaces with holes, with boundary conditions corresponding to D-instantons --
D-branes that are localized in spacetime and have finite action (or disk partition
sum). It was shown in \PolchinskiFQ\ that the combinatorics of summing over
configurations containing multiple disconnected disks with these boundary
conditions is such that the contribution of a single disk has to be exponentiated,
leading to a non-perturbative contribution
\eqn\oneinst{
\CA_{\rm D}\sim e^{Z_{\rm disk}}~.
}
Here $Z_{\rm disk}$ is the disk partition sum of the D-instanton; it goes
like $Z_{\rm disk}\sim 1/g_s$, as required for Shenker's arguments.

The idea that D-instantons provide an important source of non-perturbative effects
in string theory proved to be very fruitful in many contexts, but remarkably the
original setting in which these effects were first discussed, $c\le 1$ string
theory, remained mysterious for a long time.\foot{But, see \refs{\FukumaTJ,\NevesXT}
for some discussions.} The main problem was identifying the localized branes
that could lead to such effects. Indeed, if one naively tries
to construct a localized D-brane in Liouville theory, since the energy of the
brane goes like $1/g_s$, and the string coupling depends on the Liouville field
$\phi$, $g_s\simeq \exp(Q\phi)$, it seems that the D-brane will feel a force
pushing it to the strong coupling region $\phi\to\infty$, where it becomes light,
and leads to breakdown of perturbation theory. On general grounds one would expect
the Liouville potential, $\mu\exp(2b\phi)$ to regularize the problem, but no explicit
construction of localized D-branes with sensible properties was known.

Progress came in a beautiful paper by A. and Al. Zamolodchikov \ZamolodchikovAH, who constructed
the necessary D-branes in Liouville field theory, and analyzed some of their properties.
They showed that these branes correspond to a boundary state obtained by quantization
of a classical solution for which $\phi\to\infty$ on the boundary of the world sheet.
Nevertheless, correlation functions of bulk and boundary operators on the disk with these
boundary conditions are sensible and well behaved.

\lref\McGreevyEP{
J.~McGreevy, J.~Teschner and H.~Verlinde,
``Classical and quantum D-branes in 2D string theory,''
arXiv:hep-th/0305194.
}

The open string spectrum corresponding to these D-branes is particularly simple.
For the simplest brane, labeled as the $(1,1)$ brane in \ZamolodchikovAH, one finds in the
open string sector only the conformal block of the identity. Thus, these branes are
very natural candidates for the role of D-instantons in non-critical string theory.
Indeed, it was recently proposed \refs{\KlebanovKM,\MARTINEC} that the branes of
\ZamolodchikovAH\ are the source of the leading non-perturbative effects observed in the
matrix model.

This assertion is one of the consequences of the recent progress on the relation
between the $c\le 1$ matrix model and  two dimensional string theory
\refs{\McGreevyKB,\KlebanovKM,\McGreevyEP} , according to which
the D-branes of \ZamolodchikovAH\ are nothing but the eigenvalues of the matrices
which figure in the analysis of \refs{\DouglasVE,\BrezinRB,\GrossVS} and the relation
between the matrix model and Liouville theory is an example of an open-closed large
$N$ duality. In the matrix model, it is known that the leading non-perturbative effects are
associated with eigenvalue tunneling (see \eg\ \refs{\DavidSK,\DavidZA}); hence it is natural
that in the continuum formalism, these effects should be due to the D-branes of \ZamolodchikovAH.

The main purpose of this paper is to study the leading non-perturbative effects in the
matrix model solutions of various $c\le 1$ models coupled to Liouville and, when possible,
to compare them to the effects of D-instantons in the continuum formalism,
obtained by using the D-branes of \ZamolodchikovAH.

For $c<1$, the matrix model analysis was done in \refs{\DavidSK,\DavidZA,\EynardSG}. We construct
the appropriate D-instantons in the continuum approach, and show that they give the same leading
non-perturbative contributions as those found in \EynardSG\ (for the unitary minimal models).
This case was discussed in \MARTINEC; our results do not agree with those of \MARTINEC, since
we use a different set of Liouville $\times$ matter branes to construct the D-instantons.

For $c=1$, we study a compact scalar field $x\sim x+2\pi R$, with a winding mode perturbation
$\delta\CL=\lambda\cos R(x_L-x_R)$, coupled to world sheet gravity. We derive the leading
non-perturbative effects as a function of $\lambda$ from the matrix model, and compare a small
subset of these results to those obtained by using the branes of \ZamolodchikovAH. In all cases where
a comparison can be made, the two approaches agree. We also use the $c=1$ results to obtain
matrix model predictions for non-perturbative effects in Sine-Liouville and the two dimensional
Euclidean black hole backgrounds.

The paper is organized as follows. In section 2 we discuss unitary $c<1$ minimal models
coupled to gravity. We briefly review the matrix model analysis of \EynardSG, and then
discuss the form of the relevant D-instantons in the continuum approach. We point out
that the quantity calculated in \EynardSG\ is very natural from the continuum point of view,
compute it and show that the continuum result agrees with the matrix model one.

In section 3 we discuss the (T-dual of the) Sine-Gordon model coupled to gravity.
Using the results of \KazakovPM, we derive a differential equation that governs the non-perturbative
effects as a function of the Sine-Gordon coupling $\lambda$. We solve this equation, and use
the solution to obtain information about the leading non-perturbative effects in Sine-Liouville theory,
and for a particular value of $R(=3/2$ of the self dual radius), for the Euclidean two dimensional
black hole (or cigar) with $c=26$. We also perform a few continuum (Liouville) calculations,
which reproduce some aspects of the matrix model analysis.

In section 4 we conclude and comment on our results. A few appendices contain useful technical
results.

\newsec{  Non-perturbative effects in unitary minimal models coupled to gravity}

\lref\DiFrancescoNK{
P.~Di Francesco, P.~Mathieu and D.~Senechal,
``Conformal Field Theory,''  Springer (1997).
}

In this section we study the leading non-perturbative effects in minimal models
coupled to gravity. We restrict the discussion
to the case of unitary minimal models, since the
matrix model analysis for this case was already done in \EynardSG. It should be straightforward
to generalize our results to the non-unitary minimal model case.

Minimal models (for a review see \eg\ \DiFrancescoNK) are labeled by two integers $(p,p')$.
The unitary case corresponds to $p'=p+1$. The central charge of the $(p,p+1)$ model is given by
\eqn\cunit{
c=1-{6\over p(p+1)}~.
}
The minimal model contains a finite number of primaries of the Virasoro algebra labeled
by two integers, $(m,n)$, whose dimensions are
\eqn\dimmn{
h_{m,n}={[(p+1)m-pn]^2-1\over 4p(p+1)}~,
}
where the labels $(m,n)$ run over the range $m=1,2,\dots, p-1$,
$n=1,2,\dots, p$, and one identifies the pairs $(m,n)$ and $(p-m,p+1-n)$.

The coupling to gravity leads in conformal gauge to the appearance of the Liouville
field, $\phi$, which is governed by the action
\eqn\LIOU{
S_L= \int {d^2\s\over 4\pi} \left( (\p\phi )^2 + Q\hat R\phi
+\mul e^{2b \phi} \right)~.
}
The central charge of the Liouville model is given by
\eqn\cliouv{c=1+6Q^2~}
and the parameter $b$ is related to Q via the relation
\eqn\bQ{
Q=b+{1\over b}~.
}
In general, $b$ and $Q$ are determined by the requirement that the total central charge
of matter and Liouville is equal to $26$. In our case, \cunit\ and \cliouv\
imply that
\eqn\bminmod{
b=\sqrt{p\over p+1}~.
}
An important class of conformal primaries in Liouville theory corresponds to the
operators
\eqn\opalph{V_\alpha(\phi)=e^{2\alpha\phi}}
whose scaling dimension is given by $\Delta_\alpha=\bar\Delta_\alpha=\alpha(Q-\alpha)$.
The Liouville interaction in \LIOU\ is $\delta\CL=\mul V_b$.

The partition sum and correlation functions of the minimal models \cunit\
coupled to gravity were computed using matrix models. We start this section
by briefly reviewing the analysis of the leading non-perturbative effects in these models
due to \EynardSG. In the next subsection we will discuss the D-instantons that give
rise to these effects.

\subsec{Matrix model results}

We start with the simplest case,
$q=2$, corresponding to pure gravity. The partition sum of the model
is given by the solution of the Painleve-I equation for the second
derivative of the partition function $F(\mu)$, $u(\mu)=-\p_\mu^2 F(\mu)$:
\eqn\PAIN{
u^2(\mu)-{1\over 6} u''(\mu)=\mu~. }
We denoted the cosmological constant in the matrix model by $\mu$; it differs
from the Liouville cosmological constant in \LIOU, $\mul$, by a multiplicative
factor.

String perturbation theory is in this case an expansion in even powers of
$g_s=\mu^{-5/4}$:
\eqn\PAINs{
u(\mu)=\mu^{1/2}\sum_{h=0}^\infty c_h \mu^{-5h/2}~,  }
where $c_0=1,c_1= -1/48,\ldots$, and $c_h\under{\sim}{h\to\infty}- a^{2h}\G(2h-1/2)$,
with $a=5/8\sqrt3$.
The series \PAINs\ is asymptotic, and hence non-perturbatively ambiguous. The
size of the leading non-perturbative ambiguities can be estimated as follows. Suppose
$u$ and $\tilde u$ are two solutions of \PAIN\ which share the asymptotic behavior
\PAINs. Then, the difference between them, $\eps=\tilde u-u$, is exponentially small
in the limit $\mu\to\infty$, and we can treat it perturbatively. Plugging $\tilde u=u+\eps$
into \PAIN, and expanding to first order in $\eps$, we find that
\eqn\epsueps{
\eps''=12u\eps
}
which can be written for large $\mu$ as
\eqn\REPS{
{\eps'\over\eps}=r\sqrt{u}+b{u'\over u}+\cdots
}
with $r=-2\sqrt3$, $b=-1/4$. Using \PAINs, $u=\sqrt\mu+\cdots$, one finds that
\eqn\EXPCOR{
\eps\propto\mu^{-{1\over8}}e^{-{8\sqrt{3}\over5}\mu^{5\over4}}~.
}
The constant of proportionality in \EXPCOR\ is a free parameter of the solution,
and cannot be determined solely from the string equation \PAIN\ without further
physical input.

The authors of \EynardSG\ generalized the analysis above to the case of
$(p,p+1)$ minimal models \cunit\ coupled to gravity. They found it convenient
to parameterize the results in the same way as in the $c=0$ case, \REPS, \ie\
to define the quantity $r$,
\eqn\rdeff{{
\eps'\over\eps}=r\sqrt{u}+\cdots~,
}
where $\eps$ is again the leading non-perturbative ambiguity
in $u=-F''$. They found that for general $p$ there is in fact
a whole sequence of different solutions for $r$ labeled by two integers
$(m,n)$ which vary over the same range as the Kac indices in eq. \dimmn.
The result for $r_{m,n}$ was found to be:
\eqn\MMR{
r_{m,n}=-4\sin{\pi m\over p}\sin{\pi n\over p+1}~. }
Our main purpose in the rest of this section is to reproduce the
result \MMR\ from Liouville theory.

\subsec{Liouville analysis}

As discussed in the introduction, general considerations
suggest that the leading non-perturbative effects
in string theory should be due to
contributions of world sheets with holes, with boundary
conditions corresponding to localized D-branes.
The first question that we need to address is which D-branes
should be considered for this analysis.

The minimal model part of the background can be thought
of as a finite collection of points. All D-branes
corresponding to it are localized and therefore should
contribute to the non-perturbative effects. Minimal model
D-branes are well understood \CARDY. They are in one to one
correspondence with primaries of the Virasoro algebra
\dimmn. For our purposes, the main property that will be important
is the disk partition sum (or boundary entropy) corresponding
to the $(m,n)$ brane, which is given by
\eqn{\cormn}{Z_{m,n}=\({8 \over p(p+1)}\)^{1/4}
{\sin{\pi m \over p}\sin{\pi n \over p+1} \over
\( \sin{\pi  \over p}\sin{\pi  \over p+1}\)^{1/2}}~. }
What about the Liouville part of the background? The authors of \ZamolodchikovAH\
introduced an infinite sequence of localized D-branes, labeled by
two integers $(m',n')$. Which of these branes should we take in
evaluating instanton effects?

The analysis of \ZamolodchikovAH\ shows that open strings
stretched between the $(m',n')$ and $(m'',n'')$
Liouville branes belong to one of a finite number
of degenerate representations of the Virasoro algebra
with central charge \cliouv. The precise set of degenerate
representations that arises depends on $m',n',m'',n''$.
Degenerate representations at $c>25$ occur at negative values of
world sheet scaling dimension, except for the simplest degenerate operator,
$1$, whose dimension is zero. One finds \ZamolodchikovAH\ that in all sectors
of open strings, except those corresponding to $m'=n'=m''=n''=1$
there are negative dimension operators. It is thus natural to conjecture
that the only stable D-instantons correspond to the case $(m',n')=(1,1)$,
and we will assume this in the analysis below.

To recapitulate, the D-instantons that give rise to the non-perturbative
effects in $c<1$ minimal models coupled to gravity have the form:
$(1,1)$ brane in Liouville $\times$ $(m,n)$ brane in the minimal model.
We next show that these D-branes give rise to the correct leading
non-perturbative effects \MMR.

In order to compare to the matrix model results we should in principle
use eq. \oneinst, and evaluate the disk partition sum of the Liouville
$\times$ minimal model D-brane. It turns out, however, much more convenient
to compare directly the quantity $r$ \rdeff\ which appears naturally in
the matrix model analysis. The basic point is that this quantity is a
natural object to consider in the continuum approach as well.

Indeed, from the continuum point of view, $r$ is the ratio
\eqn{\CFTr}{
r={{\p\over \p\mu}\log \eps \over \sqrt{u}}=
{{\p \over \p\mul} Z_{\rm disc}\over \sqrt{-{\p_{\mul}^2 F}}}~, }
where in the numerator we used the fact that $\log \eps$ is the disk
partition sum corresponding to the D-instanton (see \oneinst).
Thus we see that $r$ is the ratio between the one point function of the cosmological
constant operator $V_b$ on the disk, and the square root of its two point function on the
sphere. This is a very natural object to consider since it is known in general in CFT
that $n$ point functions on the disk behave like the square roots
of $2n$ point functions on the sphere. In particular, for the purpose of computing
$r$, we do not have to worry about the multiplicative factor relating $\mu$ and $\mul$,
as it drops out in the ratio; $r$ is a pure number.
We next compute it using the results of \ZamolodchikovAH.

We start with the numerator in \CFTr. We have
\eqn\DISKP{
{\p\over \p\mul}Z_{\rm disk}= Z_{m,n} \times \langle V_b\rangle_{(1,1)}~,
}
where we used the fact that the contribution of the minimal model is simply the
disk partition sum \cormn, and the second factor is the one point function of
the cosmological constant operator \LIOU\ on the disk with the boundary conditions
corresponding to the $(1,1)$ D-brane of \ZamolodchikovAH.

$Z_{m,n}$ is given by eq. \cormn. The one point function
of $V_b$ can be computed as follows.\foot{We are grateful
to Al. Zamolodchikov for very useful discussions of this issue.}
The annulus partition sum corresponding to open strings ending
on the $(1,1)$ brane of \ZamolodchikovAH\ is given in the open string
channel by
\eqn\annone{
Z_{1,1}(t)={q^{-{Q^2\over 4}}(1-q)\over \eta(q)}~,
}
where $q=\exp(-2\pi t)$ is the modulus of the annulus, and
$\eta(q)=q^{1\over 24}\prod_{n=1}^\infty(1-q^n)$ is the
Dedekind eta function. Performing the modular transformation
to the closed string channel, $t'=1/t$, one finds \ZamolodchikovAH:
\eqn\defwave{Z_{1,1}(t')=\int_{-\infty}^\infty dP
\Psi_{1,1}(P) \Psi_{1,1}(-P)\chi_P(q')~,}
where $\Psi_{1,1}(P)$ is given in appendix A, $q'=\exp(-2\pi t')=\exp(-2\pi/t)$,
and
\eqn\charcl{
\chi_P(q)={q^{P^2}\over \eta(q)}
}
is a non-degenerate Virasoro character. $\Psi_{1,1}(P)$
can be interpreted as an overlap between the $(1,1)$ boundary state,
$B_{1,1}$, and the state with Liouville momentum $P$,
\eqn\psioverlap{
\Psi_{1,1}(P)=\langle B_{1,1} |P\rangle~.
}
Therefore, $\Psi_{1,1}(P)$ is proportional to the one point function
on the disk, with $(1,1)$ boundary conditions, of the Liouville operator
$V_\alpha$, with
\eqn\alphaP{
\alpha={Q\over2}+iP~.
}
The proportionality constant is a pure number (independent of $P$ and $Q$).
We will not attempt to calculate this number precisely here, since we can deduce
it by matching any one of the matrix model predictions, and then use it in all
other calculations, but we will next mention a few contributions to it, for illustrative
purposes.

First, in \defwave\ the $P$ integral runs over the whole real line, but $|P\rangle$
and $|-P\rangle$ are the same state, due to reflection from the Liouville wall,
so we should replace $\int_{-\infty}^\infty dP =2 \int_{0}^\infty dP $. Hence, the
physical wavefunctions which should appear in \psioverlap\ are $\sqrt{2}\Psi_{1,1}$.
Furthermore, the state $|P\rangle$ in \psioverlap\ is normalized as
$\langle P|P'\rangle=\delta(P-P')$, for consistency with \defwave. On the other hand,
the state created by acting with the Liouville exponential $V_\alpha$ \opalph, with
$\alpha$ given by \alphaP, is normalized to  $\pi\delta(P-P')$. Finally, when
computing $\langle V_\alpha(0)\rangle_{\rm disk}$, the part of the $SL(2,\IR)$
symmetry of the disk related to rotations around the origin is unfixed. This gives
an extra factor of $1/2\pi$ in the one point function. Altogether, one finds
\eqn\oneptnorm{\langle V_\alpha(0)\rangle_{(1,1)}=C\cdot\sqrt{2}\cdot\sqrt{\pi}\cdot
{1\over 2\pi}\cdot \Psi_{1,1}}
where $\alpha$ is given by \alphaP\ and $C$ is an undetermined constant
which we will fix by comparing to matrix model results (we will find that
$C=2$).

We can now plug in \oneptnorm\ into \DISKP\ and, using the form of $\Psi_{1,1}$
given in (A3), find
\eqn{\wavefP}{
\langle V_b(0)\rangle_{(1,1)}=
-C \, { 2^{1/4} \sqrt{\pi}  [\pi\mul \gamma(b^2)]^{\hf(1/b^2-1)}
\over b \Gamma(1-b^2)\Gamma(1/b^2)}~.
}
We next move on to the denominator of \CFTr. This is given by the two point function
$\langle V_bV_b\rangle_{\rm sphere}$. This quantity was calculated in
\refs{\DornXN,\ZamolodchikovAA}. It is convenient
to first compute the three point function  $\langle V_bV_bV_b\rangle_{\rm sphere}$ and then
integrate once, to avoid certain subtle questions regarding the fixing of the $SL(2,\IC)$
Conformal Killing Group of the sphere. One has (see appendix A):
\eqn{\threebb}{\langle V_bV_bV_b\rangle_{\rm sphere}=
b^{-1}\left[\pi\mul
\right]^{1/b^2-2}[\gamma(b^2)]^{1/b^2}\gamma(2-1/b^2)~. }
Integrating once w.r.t. $-\mul$ we find
\eqn{\twob}{\langle V_bV_b\rangle_{\rm sphere}=
{1/b^2-1\over \pi b} \left[\pi\mul\gamma(b^2)\right]^{1/b^2-1}
\gamma(b^2)\gamma(1-1/b^2)~.
}
We are now ready to compute $r$, \CFTr. Plugging in \cormn, \wavefP\ and
\twob\ into \CFTr, we find
\eqn\RFIN{
r_{m,n}=-2C\sin{\pi m \over p}\sin{\pi n \over p+1}~,
}
which agrees with the matrix model result \MMR\ if we set $C=2$.
Since $C$ is independent of $m$, $n$ and $p$,
we can fix it by matching to any one
case, and then use it in all others. We conclude that the Liouville analysis
gives the same results as the matrix model one.

\newsec{Non-perturbative effects in compactified $c=1$ string theory}

In this section we discuss the leading non-perturbative effects
in (Euclidean) two dimensional string theory. The world sheet
description contains a field $x$, compactified on a circle of
radius $R$, $x\simeq x+2\pi R$, coupled to gravity. The conformal
gauge Lagrangian describing this system is
\eqn\conepert{
\CL={1\over 4\pi}\left[(\partial x)^2 +(\partial\phi)^2
+2\hat R\phi+\mul \phi e^{2\phi}+\lambda e^{(2-R)\phi}
\cos [R(x_L-x_R)]\right]~.
}
where we chose $\alpha'=1$, and perturbed the $c=1$ model by a
Sine-Gordon type interaction, $\lambda \cos [R(x_L-x_R)]$, which gives
rise to the last term in \conepert\ after coupling to gravity. The
perturbation is relevant for $R<2$, and we will restrict to that case
in the discussion below.

We will next describe the matrix model predictions for the leading
non-perturbative effects corresponding to \conepert, and then verify
some of them in Liouville theory.

\subsec{Matrix model analysis}

The model \conepert\ was solved using matrix model techniques in \KazakovPM,
extending some earlier results \refs{\MooreGA,\DijkgraafHK,\HsuCM,\BoulatovXZ,
\HoppeXG}.
In particular, it was shown in \KazakovPM\ that the Legendre transform $\FSL$ 
of the string partition sum $F$ satisfies the Toda differential equation:
\eqn{\todafd}{
{1\over 4}\lambda^{-1}\p_{\lambda}\lambda\p_{\lambda} \FSL(\mu, \lambda)+
\exp\left[-4\sin^2\(\hf{\p\over \p\mu}\) \FSL(\mu,\lambda)\right]=1~. }
The initial condition for this equation is supplied by the 
unperturbed $c=1$ string theory on a circle
\eqn\FrenO{\eqalign{
&\hskip 3.3cm
\FSL(\mu,0) =
{R\over 4} \Re \int_{\Lambda^{-1}}^\infty {ds\over s}
{ e^{-i\mu s}\over \sinh{s\over 2}\sinh{s\over 2R}} \cr
&\hskip -0.3cm
 = -  {R\over2}\mu^2 \log {\mu\over\Lambda} -
{1\over24}\big(R + {1\over R} \big)\log {\mu\over\Lambda} +
R\sum_{h=2}^\infty \mu^{2-2h} c_h(R)+O(e^{-2\pi\mu})+O(e^{-2\pi R \mu})~,
}}
where the genus $h$ term $c_h(R)$ is a known polynomial in ${1/ R}$. 
The function $\FSL(\lambda,\mu)$ has a genus expansion which follows
from \todafd. The string partition sum $F$ is obtained from $\FSL$ by
flipping the sign of the genus zero term \KLEBANOV. 

We will need below some features of the solution of equations
\todafd, \FrenO, which we review next. In order to study the 
solution of \todafd, \FrenO, it is convenient to introduce the 
parameters
\eqn{\scp}{
y=\mu\xi, \qquad \xi=(\lambda\sqrt{R-1})^{-{2\over 2-R}}~.
}
This parameterization is useful when $\lambda$ is large,
so that the last term in \conepert\ sets the scale,
and one can study the cosmological term perturbatively in $\mu$.
As explained in \KazakovPM, this can only be done for $R>1$. This is the
physical origin of the branch cut in the definition of $\xi$ in \scp.
In the regime where $\lambda$ sets the scale, the genus expansion
of the string partition sum is an expansion in even powers of $\xi$:
\eqn{\gnexpCF}{
\FSL(\mu,\lambda)=\lambda^2+
\xi^{-2}\left[{R\over 2}y^2\log\xi+ f_0(y)\right]
+\left[ 
{R+R^{-1} \over 24}\log\xi+f_1(y)\right]+
\sum\limits_{h=2}^{\infty}\xi^{2h-2}f_h(y)~. }
The partition sum on the sphere, $\FSL_0(\mu,\l)$,
is given by the second term on the r.h.s. of \gnexpCF.
One can show \KazakovPM\ that at this order, the full Toda equation
\todafd\ reduces to an algebraic equation for $\XX=\p^2_y f_0$
\eqn{\wf}{ y=\efo-\ego~.  }
This leads to the following perturbative
expansion of $\FSL_0(\mu,\l)$ in powers of $\lambda$:
\eqn{\cMoore}{
\FSL_0(\mu,\lambda)={R\over 2}\mu^2\log{\Lambda\over\mu}
+R\mu^2\sum\limits_{n=1}^{\infty}{1\over n!} \((1-R)\mu^{R-2}\lambda^2\)^n
{ \Gamma(n(2-R)-2)\over \Gamma(n(1-R)+1)}~,
}
which agrees with the original (conjectured) result for this
quantity in \MooreGA. The genus $h\ge 1$ terms in the expansion
\gnexpCF\ can be computed in principle by substituting \gnexpCF\ into
\todafd, but only $f_0$ and $f_1$ are known in closed form. Nevertheless,
we will see below that the large $h$ behavior of $f_h$ can be obtained
directly from the differential equation \todafd, by computing the
leading non-perturbative corrections to \gnexpCF. We next turn to
these corrections.

At $\lambda=0$,
there are two types of non-perturbative corrections to the $1/\mu$
expansion \FrenO, associated with the poles of the integrand
in that equation. These occur at $s=2\pi ik$ and
$s=2\pi R ik$, $k\in \IZ$, and give rise to the non-perturbative
effects indicated on the second line of \FrenO, $\exp(-2\pi\mu k)$ and
$\exp(-2\pi R\mu k)$, respectively.

At finite $\lambda$, the situation is more interesting. The series of
non-perturbative corrections
\eqn\dirnonpert{
\Delta \FSL=\sum_{n=1}^\infty C_n  e^{ -2\pi n\mu}
}
gives rise to an exact solution of the {\it full} equation \todafd\
\KazakovPM. Thus, the corresponding instantons are
insensitive to the presence of the sine-Gordon
perturbation! We will return to this interesting fact, and explain
its interpretation in Liouville theory, in the next subsection.

The second type of corrections, which starts at $\lambda=0$ like
$\Delta \FSL=e^{ -2\pi R\mu} $, does not solve the full equation \todafd,
and gets $\lambda$ dependent corrections. To study these corrections
one can proceed in a similar way to that described in section {\it 2.1}.

Let $\FSL$ and $\tilde \FSL=\FSL+\eps$ be two solutions of the Toda equation
\todafd. The linearized equation for $\eps$ reads
\eqn{\todasph}{
{1\over 4}\lambda^{-1}\p_{\lambda}\lambda\p_{\lambda}\eps(\mu, \lambda)
-4e^{-\p^2_{\mu}\FSL_0(\mu,\lambda)}
\sin^2\(\hf{\p\over \p\mu}\)\eps(\mu,\lambda)=0~, }
where in the exponential in the second term we approximated
\eqn\sphfr{
4\sin^2\(\hf{\p\over \p\mu}\)\FSL(\mu,\lambda)\simeq
\p^2_{\mu}\FSL_0= R\log\xi+\XX(y)~. }
This is similar to the fact that in the discussion of section 2,
one can replace $u$ in eq. \epsueps\ by its limit as $\mu\to\infty$.
After the  change of variables from $(\lambda, \mu)$ to $(y,\mu)$,
equation \todasph\ can be written as
\eqn{\eqep}{
\alpha {y^2\over \mu^2}(y\p_y)^2\eps(\mu,y)
-4 e^{-\XX(y)}
\sin^2\(\hf\({\p\over \p\mu}+{y\over \mu}{\p\over \p y}\)\)\eps(\mu,y)=0~,
}
where $\alpha\equiv {R-1\over (2-R)^2}$.

Since we know that $\eps$ is non-perturbatively small in the
$\mu\to \infty$ limit, we use the following ansatz for it:
\eqn{\anz}{
\eps(\mu,y)= P(\mu,y) e^{-\mu f(y)}~. }
Here $P$ is a power-like prefactor in $g_s$, and $f(y)$ is the
function we are interested in (the analogue of $r$ in the minimal models
of section 2).
Substituting {\anz} into {\eqep} and keeping only the leading terms in
the $\mu\to\infty$  limit, one finds the following first order  differential
equation:
\eqn{\eqg}{
\sqrt{\alpha}\,e^{\hf \XX(y)} (1- y\p_y) g(y)
=\pm \sin\[\p_y g(y)\]~,  }
where we introduced
\eqn\ggffyy{g(y)=\hf yf(y)~.}
The $\pm$ in \eqg\ is due to the fact that one in fact finds
the square of this equation.

Equation \eqg\ is a first order differential equation in $y$, and to
solve it we need to specify boundary conditions. As discussed earlier
for the perturbative series, it is natural to specify these boundary
conditions at $\lambda\to 0$, or $y\to\infty$ (see \scp). We saw
in the discussion of \FrenO\ that there are two solutions,
$f(y\to\infty)\to 2\pi$ or $2\pi R$. This implies via \ggffyy\ that
$g(y\to\infty)\simeq \pi y$ or $\pi Ry$. We already saw that
$g(y)=\pi y$ gives an exact solution of \todafd, and this is true
for \eqg\ as well (as it should be). Thus, to study non-trivial
non-perturbative effects, we must take the other boundary condition,
\eqn{\incon}{
g(y\to\infty)\simeq\pi R y~. }

Interestingly, eq. \eqg\ is exactly solvable. We outline
the solution in Appendix B. For the initial condition \incon,
the solution can be written as
\eqn{\gphi}{
g(y)=y\phi(y) \pm {1\over \sqrt{\alpha}}e^{-\hf \XX(y)}\sin \phi(y)~,
}
where $\phi(y)=\partial_y g$ satisfies the equation
\eqn{\zzz}{
e^{{2-R\over 2R}X(y)}=\pm \sqrt{R-1}\, { \sin \({1\over R} \phi\)
\over \sin \( {R-1\over R} \phi\) }~.  }
The solution with the minus sign in \zzz\ can be shown to be unphysical
(see below). We will thus use the solution with a plus sign.

Equations \gphi, \zzz\ are the main result of this subsection.
We next discuss some features of the corresponding non-perturbative
effects.

Consider first the situation for small $\lambda$, or large $y$.
The first three terms in the expansion of $\phi(y)$ are
\eqn{\corr}{
\phi(y) \approx
\pi R\pm{R \sin(\pi R)\over \sqrt{R-1}}\, y^{-{2-R\over 2}}
+{R\over 2}\sin(2\pi R)\, y^{-(2-R)}~. }
This gives the following result for $f(y)$ \anz:
\eqn{\corrf}{\eqalign{
f(y)&=
2\pi R \pm {4\sin(\pi R)\over\sqrt{R-1}}\, y^{-{2-R\over 2}}
+{R\sin (2\pi R)\over R-1} y^{-(2-R)} +O(y^{-3(2-R)/2})
\cr
&=2\pi R \pm {4\sin(\pi R)}\, \mu^{-{2-R\over 2}}\, \lambda +
R\sin(2\pi R)\, \mu^{-(2-R)}\,\lambda^2+O(\lambda^3)~.
}}
We see that for large $y$, the expansion parameter is $y^{-{(2-R)\over2}}\sim
\lambda$, as one would expect.

Another interesting limit is  $\mu\to 0$ at fixed $\lambda$, \ie\
$y\to 0$, which leads to the Sine-Liouville model, \conepert\ with $\mul=0$.
For $R=3/2$, this model is equivalent to the Euclidean black hole
background \refs{\FZZ,\KazakovPM}. As we see in \wf, in this limit $X\to 0$.
The first two terms in the expansion of $\phi$ around this point are
\eqn\bhphi{
\phi(y)=\phi_0+
{R\over 2}
\((R-1)\cot\({R-1\over R}\phi_0\)-\cot\({1\over R}\phi_0\)\)^{-1}y +O(y^2)~,
}
where $\phi_0$ is defined by the equation
\eqn{\zzzbh}{
 { \sin \({1\over R} \phi_0\)
\over \sin \( {R-1\over R} \phi_0\) }=\pm {1\over \sqrt{R-1}}~.  }
The function $f(y)$ is given in this limit by the expansion
\eqn{\gphibh}{
f(y)= \pm {2(2-R)\over y\sqrt{R-1}}\sin \phi_0
+ 2\phi_0 + O(y)~.
}
Note that the behavior of $f$ as $y\to 0$, $f\sim 1/y$, leads
to a smooth limit as $\mu\to 0$ at fixed $\lambda$. The
non-perturbative effect \anz\ goes like $\exp(-\mu f(y))$, so
that as $y\to 0$ the argument of the exponential goes like
$\mu/y=1/\xi$, and all dependence on $\mu$ disappears.

For $R=3/2$, which corresponds to the Euclidean black hole,
the equations simplify. One can explicitly find $\phi_0$
because {\zzzbh} gives
\eqn{\eqbhR}{
 \cos {\phi_0\over 3}
= \pm {1\over\sqrt{2}} \Rightarrow \phi_0={3\pi\over 4}\ {\rm or\ }
\phi_0={9\pi\over 4}~.  }
As a result, one finds at this value of the radius a simple result
\eqn{\gbh}{
\mu f(y)= \pm {\mu \over y}+{(6\mp3)\pi\over 2}\mu+\cdots
=\pm \hf \lambda^{4}+{(6\mp3)\pi\over 2}\mu+\cdots~. }
Note that the solution with the minus sign leads to
a growing exponential, $e^{\hf \lambda^4}$.
Therefore, it is not physical, as mentioned above.
For the particular case $R=3/2$ one can write the solution for $f(y)$
more explicitly. The result is given in Appendix B.

Another nice consistency check on our solution is the RG flow from $c=1$
to $c=0$ CFT coupled to gravity, implicit in \conepert. Before
coupling to gravity, the Sine-Gordon model associated to \conepert\
describes the following RG flow. In the UV, the Sine-Gordon coupling
effectively goes to zero, and one approaches the standard CFT of a compact
scalar field. In the IR, the potential given by the Sine-Gordon interaction
gives a world sheet mass to $x$, and the model approaches a trivial $c=0$
fixed point. As shown in \refs{\MooreGA, \HsuCM}, this RG flow manifests
itself after coupling to gravity in the dependence of the physics on $\mu$.
Large $\mu$ corresponds to the UV limit; in it, all correlators approach
those of the $c=1$ theory coupled to gravity. Decreasing $\mu$ corresponds
in this language to the flow to the IR, with the $c=0$ behavior recovered
as $\mu$ approaches a critical value $\mu_c$, which is given by:
\eqn\crpoint{
\mu_c=-(2-R)(R-1)^{R\over 2-R}\lambda^{2\over 2-R}~.
}
The non-perturbative contributions to the partition sum computed in this
section must follow a similar pattern. In particular, $f(y)$ must exhibit
a singularity as $y\to y_c$, with
\eqn\yyccrr{y_c=-(2-R)(R-1)^{R-1\over 2-R}}
and furthermore, the behavior of $f$ near this singularity should reproduce
the non-perturbative effects of the $c=0$ model coupled to gravity discussed
in section 2.

The singularity at $y=y_c$ corresponds to a critical point of \wf,
near which the relation between $y$ and $X$ degenerates:
\eqn\yXrel{
{y_c-y\over y_c}\simeq {R-1\over 2R^2}(X-X_c)^2+O\((X-X_c)^3\)~.
}
Solving for the critical point, one finds that
\eqn\xxccrr{
e^{-{2-R\over 2R}X_c}=\sqrt{R-1}~.
}
The corresponding $y_c$ is indeed given by \yyccrr.
We can now solve for $f(y)$, near $y=y_c$. Substituting
\xxccrr\ in \zzz\ we find
\eqn{\zzzww}{
{ \sin \({1\over R} \phi\)\over \sin \( {R-1\over R} \phi\)}
={1\over R-1}~.
}
Thus, the $c=0$ critical point corresponds
to $\phi\to 0$.\foot{Note that if we chose
the minus sign in \zzz, we would find a more complicated solution
for $\phi$. One can show that it would lead to the wrong critical
behavior. This is an additional check of the fact that the physical
solution corresponds to the plus sign in \zzz.}
The first two terms in the expansion of $\phi$ around the singularity are
\eqn\crphi{
\phi(y)=\sqrt{3}(X_c-X)^{1/2}-{\sqrt{3}(R^2-2R+2) \over 20 R^2}(X_c-X)^{3/2}
+O\((X_c-X)^{5/2}\)~.
}
Substituting this in  \gphi\ one finds
\eqn\gcrit{
g(y)=-y_c {2\sqrt{3}(R-1)\over 5 R^2}(X_c-X)^{5/2}+O\((X_c-X)^{5/2}\)
}
or, using \ggffyy:
\eqn\fcrit{
f(y)\approx - {8\sqrt{3}\over 5 }\({2 R^2\over R-1}\)^{1/4}
\({\mu_c-\mu \over \mu_c}\)^{5/4}~.
}
The power of $\mu-\mu_c$ is precisely right to describe a leading
non-perturbative effect in pure gravity. It is interesting to
compare the coefficient in \fcrit\ to what is expected in pure gravity.
It is most convenient to do this by again computing $r$ \REPS.
$u$ is computed by evaluating the leading singular term as $\mu\to\mu_c$
in $\p^2_{\mu}\FSL_0=R\log\chi+X(y)$. One finds
\eqn\rcr{
r=-2\sqrt{3}\({2 R^2\over R-1}\)^{1/4}
\({\mu_c-\mu \over \mu_c}\)^{1/4} ( X_c-X)^{-1/2}=
-2\sqrt{3}
}
in agreement with the result \MMR\ for pure gravity. This provides another
non-trivial consistency check of our solution.

\subsec{Liouville analysis}

In this subsection we will discuss the Liouville interpretation of the
matrix model results presented in the previous subsection. Consider
first the unperturbed $c=1$ theory corresponding to $\lambda=0$ in
\conepert. As we saw, in the matrix model analysis one finds two
different types of leading non-perturbative effects (see \FrenO),
$\exp(-2\pi\mu)$, and $\exp(-2\pi R\mu)$. It is not difficult to guess
the origin of these non-perturbative effects from the Liouville point of
view. The $\exp(-2\pi\mu)$ contribution is due to a D-brane that corresponds
to a $(1,1)$ Liouville brane $\times$ a Dirichlet brane in the $c=1$ CFT,
\ie\ a brane located at a point on the circle parameterized by $x$.
Similarly, the $\exp(-2\pi R\mu)$ term comes from a brane wrapped around
the $x$ circle.

This identification can be verified in the same way as we did for
minimal models in section 2. To avoid normalization issues, one can
again calculate the quantity $r$ \CFTr. The matrix model prediction
for the Neumann brane\foot{Similar formulae can be written for the
Dirichlet brane.} is
\eqn\rmu{
r=-{2\pi \sqrt{R} \over \sqrt{\log{\Lambda\over\mu}}}~,
}
where we used the sphere partition function $\FSL_0(\mu,0)$ of the
unperturbed compactified $c=1$ string given by the first term in
\cMoore.

The CFT calculation is similar to that performed in the $c<1$ case.
The partition function of $c=1$ CFT on a disk with Neumann boundary
conditions is well known; the calculation is reviewed in Appendix C.
One finds
\eqn{\pfdisc}{
Z_{\rm Neumann}=2^{-1/4}\sqrt{R}~. }
The disk amplitude corresponding to the $(1,1)$ Liouville brane,
and the two-point function on the sphere are computed using equations
\wavefP\ and \twob, in the limit $b\to 1$. The limit is actually singular,
but computing everything for generic $b$ and taking the limit at the end
of the calculation leads to sensible, finite results. The leading behavior
of \wavefP\ as $b\to 1$ is
\eqn\psii{\langle V_b(0)\rangle_{(1,1)}\approx - {2^{5/4}\sqrt{\pi}\over \Gamma(1-b^2)}~,
}
with the constant $C$ in \wavefP\ chosen to be equal to $2$, as in the minimal model
analysis of equations \oneptnorm\ -- \RFIN.
The two point function \twob\ approaches
\eqn{\twomu}{
\p^2_{\mul} \FSL_0\simeq
-{ \log\mul\over \pi\Gamma^{2}(1-b^2)}~.}
Substituting {\pfdisc}, {\psii} and {\twomu}
into {\CFTr} gives the result {\rmu}. This provides
a non-trivial check of the statement in section 2, that
the constant $C$ in eq. \oneptnorm\ is a pure number, independent
of all the parameters of the model. 

The agreement of \rmu\ with the Liouville analysis supports
the identification of the Neumann D-branes as the source
of the non-perturbative effects $\exp(-2\pi R\mu)$. A similar analysis
leads to the same conclusion regarding the relation between the Dirichlet
$c=1$ branes and the non-perturbative effects $\exp(-2\pi\mu)$ (the two
kinds of branes are related by T-duality).

Having understood the structure of the unperturbed theory, we next turn to
the theory with generic $\lambda$. In the matrix model we found
that the non-perturbative effects associated with the Dirichlet brane localized
on the $x$ circle are in fact independent of $\lambda$ (see \dirnonpert\
and the subsequent discussion). How can we understand this statement from the
Liouville point of view?

The statement that \dirnonpert\ is an exact solution of the Toda
equation \todafd\ corresponds in the continuum formulation to the
claim that the disk partition sum of the model \conepert, with $(1,1)$
boundary conditions for Liouville, and Dirichlet boundary conditions
for the matter field $x$ is independent of $\lambda$.  In other words,
all $n$-point functions of the operator given by the last term in
\conepert\ on the disk vanish
\eqn\vanishcor{
\langl \left(\int d^2z e^{(2-R)\phi}
\cos [R(x_L-x_R)]\right)^n\rangl_{(1,1)\times{\rm Dirichlet}}=0~.
}
Is it reasonable to expect \vanishcor\ to be valid from the world
sheet point of view?  For odd $n$ \vanishcor\ is trivially zero
because of winding number conservation.  Indeed, the Dirichlet
boundary state for $x$ breaks translation invariance, but preserves
winding number. The perturbation in \conepert\ carries winding number,
and for odd $n$ all terms in \vanishcor\ have non-zero winding
number. Thus, the correlator vanishes.

For even $n$ one has to work harder, but it is still reasonable to
expect the amplitude to vanish in this case. Indeed, consider the
T-dual statement to \vanishcor, that the $n$ point functions of the
momentum mode $\cos(x/R)$, on the disk with $(1,1)$ $\times$ Neumann
boundary conditions, vanish. This is reasonable since the operator
whose correlation functions are being computed localizes $x$ at the
minima of the cosine, while the D-brane on which the string ends is
smeared over the whole circle.  It might be possible to make this
argument precise by using the fact that in this case the D-instanton
preserves a different symmetry from that preserved by the perturbed
theory, and thus it should not contribute to the non-perturbative
effects.

To summarize, the matrix model analysis predicts that \vanishcor\ is
valid. We will not attempt to prove this assertion here from the
Liouville point of view (it would be nice to verify it even for the
simplest case, $n=2$), and instead move on to discuss the
non-perturbative effects due to the localized branes on the $x$
circle.

The non-trivial solution of eq. \eqg\ given by \gphi, \zzz\ should
correspond from the Liouville point of view to the disk partition sum
of the $(1,1)$ Liouville brane which is wrapped around the $x$
circle. The prediction is that
\eqn\nonvanishcor{
\langl \left(\int d^2z e^{(2-R)\phi}
\cos [R(x_L-x_R)]\right)^n\rangl_{(1,1)\times{\rm Neumann}}
}
are the coefficients in the expansion of $f(y)$ \anz\ in a power
series in $\lambda$, the first terms of which are given by \corrf. It
would be very nice to verify this prediction directly using Liouville
theory, but in general this seems hard given the present state of the
art. A simple check that can be performed using results in
\ZamolodchikovAH\ is to compare the order $\lambda$ term in \corrf\
with the $n=1$ correlator \nonvanishcor. We next compare the two.

Like in the other cases studied earlier, it is convenient to define a
dimensionless quantity, $\rho$, given by the ratio of the one point
function on the disk \nonvanishcor\ and the square root of the
appropriate two point function on the sphere,
\eqn{\rcone}{
\rho=\left. {{\p\over \p \lambda} \log \eps
\over \sqrt{-\p_{\lambda}^2\FSL_0}}\right|_{\lambda=0}~. }
The matrix model result \corrf, \cMoore, for this quantity is
\eqn{\rconeMM}{
\rho=-\left.{\mu {\p\over \p \lambda} f \over
\sqrt{-\p_{\lambda}^2\FSL_0}}\right|_{\lambda=0}
= -2\sqrt{2}\sin(\pi R)~.  }
In the Liouville description, $\rho$ is given by
\eqn{\CFTrho}{
\rho= {B_\CT \langle V_{b-{R\over2}}\rangle_{(1,1)}
\over \sqrt{-\langle\CT^2\rangle}}~.
}
Here, $\CT=\cos R(x_L-x_R)V_{b-{R\over2}}$.
$B_{\CT}$ is the one point function of $\cos R(x_L-x_R)$ on the disk.
It has the same value as \pfdisc\ (see appendix C)
\eqn{\bt}{
B_{\CT}=2^{-1/4}\sqrt{R}~. }
The one point function of the operator  $V_{b-{R\over2}}$ is related
to the wavefunction (A.3) with momentum $iP=b-R/2-Q/2$ and is given by
\eqn{\wavefPR}{
\langle V_{b-{R\over2}}(0)\rangle_{(1,1)}=
-{ 2^{5/4} \sqrt{\pi}  [\pi\mul \gamma(b^2)]^{\hf(1/b^2-1+R/b)}
\over b \Gamma(1-b^2+Rb)\Gamma(1/b^2+R/b)}~.  }

The two-point function of $\CT$ on the sphere
is computed as above from the three point function (A.10).
One finds
\eqn{\twol}{
\langl \CT^2\rangl ={\( {1\over b^2}+{R\over b}-1\)\over 2\pi b}
\left[\pi\mul \gamma(b^2)\right]^{{1\over b^2}+{R\over b}-1}
\gamma\(b^2-Rb\) \gamma\(1-{1\over b^2}-{R\over b}\)~.
}
Substituting these results into \CFTrho\ leads,
in the limit $b=1$, to \rconeMM. We see that the 
Liouville results are again in complete agreement with
the corresponding matrix model calculation.

\newsec{ Discussion }

In this paper we studied non-perturbative effects in $c\le 1$
non-critical string theory, using both matrix model and continuum
(Liouville) methods.  We showed that the matrix model results
correspond in the continuum approach to D-instanton effects due to
localized branes in Liouville theory, the $(1,1)$ branes constructed
in \ZamolodchikovAH. Our work is motivated by the recent proposal of
\MARTINEC, but the details are different.

For $c<1$ unitary minimal models, we showed that the contribution of
these branes reproduces the matrix model results of \EynardSG. For
$c=1$ we used the results of \KazakovPM\ to derive the leading
non-perturbative effects in the matrix model description of the
Sine-Gordon model coupled to gravity, as a function of the
compactification radius and the Sine-Gordon coupling, and reproduced
some of these predictions using the Liouville branes of
\ZamolodchikovAH.

The matrix model analysis led in addition to a number of predictions
regarding the properties of D-instantons in $c=1$ string theory. In
particular, we showed that an infinite number of correlation
functions given in eq. \vanishcor\ must vanish, and the correlation functions
\nonvanishcor\ are given by the expansion of $f(y)$ \ggffyy\ in a power
series in $\lambda$, the first few terms of which appear in \corrf.

Another interesting set of matrix model predictions concerns the limit
$\mu\to0$, in which one finds the Sine-Liouville model, \conepert\
with $\mu_L=0$. The expansion of $f(y)$ around $y=0$, \gphibh,
provides information about the disk partition sum and correlation
functions of closed string operators in the presence of D-branes in
this model. For the particular case $R=3/2$, these predictions apply
also to the $SL(2)/U(1)$ (cigar) background corresponding to a
Euclidean two dimensional black hole.  In the cigar, the D-branes in
question are wrapped around the angular direction of the cigar, and
are localized near the tip. Our results give rise to predictions for
correlation functions on the disk with these boundary conditions.

Our analysis also resolves a previously open problem, of describing
world sheet gravity on the disk with the boundary conditions described
in \ZamolodchikovAH\ using the matrix model approach. It opens the possibility
of studying gravity on $AdS_2$ using matrix model techniques.

An interesting open problem is how to go beyond the calculation of the
leading non-perturbative contributions to the amplitudes \ampl, and
study the perturbative expansion around the D-instantons discussed in
this paper. In the matrix model approach one finds that the leading
non-perturbative terms behave like
\eqn\onein{
\CA_{1-inst.}\sim C\ g_s^{f_A}\ e^{-{f_D \over g_s}}~,
}
where $f_D$ was discussed in this paper, and $f_A$ is known as
well.\foot{For the partition sum one finds a simple universal
behaviour. For the minimal models, in all cases that we are aware of,
$F_{\rm non-pert}\sim g_s^{1/2}e^{-{f_D \over g_s}}$ (see
\GinspargCY).  For the $c=1$ model, equation \dirnonpert\ suggests
that $\FSL_{\rm non-pert}\sim g_s^0 e^{-{f_D \over g_s}}$.}  The
constant $C$ is ambiguous in the matrix model.

In the continuum Liouville description, $f_D$ has been computed in this
paper, and it would be interesting to compute $f_A$ and understand whether
$C$ is ambiguous or can be determined. In order to study these issues,
one has to understand the perturbative expansion about the D-instantons
discussed in this paper. In particular, it is natural to expect that $f_A$
will arise from the annulus with boundary conditions corresponding to the
$(1,1)$ brane (this gives the correct scaling with $g_s$).

The discussion of this paper has some interesting
higher dimensional generalizations. For example, if one
replaces Liouville by $N=2$ Liouville, and the minimal
models by $N=2$ minimal models, one finds the background
\eqn\slsu{{SL(2)_k\over U(1)}\times {SU(2)_k\over U(1)}}
describing $NS5$-branes spread out on a circle
\refs{\GiveonPX,\GiveonTQ}. The D-instantons analogous to those
studied in this paper correspond to Euclidean $D0$-branes stretched
between pairs of fivebranes. From the point of view of the geometry
\slsu, they are described by D-branes living near the tip of the
cigar, which are $N=2$ superconformal generalizations of the $(1,1)$
brane of \ZamolodchikovAH. In particular, the spectrum of open strings
ending on them corresponds to the $N=2$ superconformal block of the
identity.

Like in our case, the D-instantons are obtained by tensoring the above
$N=2$ Liouville branes with the different D-branes in the $N=2$
minimal model. Understanding the non-perturbative effects due to these
instantons is an interesting open problem. Studying the branes
discussed in this paper in more detail is a very useful warmup
exercise for the fivebrane problem. It might suggest hints for finding
an analog of the matrix model for fivebranes.

\bigskip
\noindent{\bf Acknowledgements:}
We thank V. Fateev, A. Giveon, I. Kostov, E. Martinec, G. Moore,
and especially Al. Zamolodchikov for useful discussions. 
The work of V.K. is supported in part by NATO grant  PST.CLG.978817.
The work of D.K. is supported in part
 by DOE grant
\#DE-FG02-90ER40560.  D.K. thanks LPT at Ecole Normale Sup\'erieure (Paris), 
LPTHE at Universit\'e Paris VI, and LPM at Universit\'e Montpellier II
for hospitality during the course of this work.  The work of S.A. and
V.K. was partially supported by European Union under the RTN contracts
HPRN-CT-2000-00122 and -00131. The work of S.A. was also supported in
part by European network EUCLID HPRN-CT-2002-00325.

\appendix{A}{ Correlation functions in Liouville theory.}

In this appendix we summarize some results on Liouville field
theory, from \refs{\ZamolodchikovAH,\DornXN,\ZamolodchikovAA}, which
are used in the text. The one-point function of the operator \opalph\
in Liouville theory on the disk with boundary conditions
corresponding to the $(1,1)$ brane is given by
\eqn{\onecorL}{
U(\alpha)={ [\pi\mul \gamma(b^2)]^{-\alpha/b} Q\Gamma(bQ)\Gamma(Q/b)
\over (Q-2\alpha)\Gamma(b(Q-2\alpha))\Gamma((Q-2\alpha)/b)}~,
}
where
\eqn\ggaamm{
\gamma(x)={\Gamma(x)\over\Gamma(1-x)}~.
}
$U(\alpha)$ is the normalized one-point function, \ie\ the one point function
divided by the disk partition function. As discussed in the text,
the unnormalized one point function of $V_\alpha$ is related to the wavefunction
with momentum $iP=\alpha-Q/2$ defined by \defwave, which is given by
\eqn{\wavef}{
\Psi_{1,1}(P)={ 2^{3/4} 2\pi i P [\pi\mul \gamma(b^2)]^{-iP/b}
\over \Gamma(1-2ibP)\Gamma(1-2iP/b)}~.
}

The three-point function on the sphere is (suppressing the
standard dependence on the world sheet positions of the
vertex operators)
\eqn{\three}{\eqalign{
&\hskip 1.5cm
\langle V_{\alpha_1}V_{\alpha_2}V_{\alpha_3}\rangle_{\rm sphere}=
\left[\pi\mul \gamma(b^2)b^{2-2b^2}\right]^{\(Q-\sum\alpha_i\)/b}\times
 \cr
&{\U_0\U(2\alpha_1)\U(2\alpha_2)\U(2\alpha_3) \over
\U(\alpha_1+\alpha_2+\alpha_3-Q)\U(\alpha_1+\alpha_2-\alpha_3)
\U(\alpha_2+\alpha_3-\alpha_1)\U(\alpha_3+\alpha_1-\alpha_2)}~,
}}
where
\eqn{\Ups}{
\U(x)=\exp\left\{ \int\limits_0^{\infty} {dt\over t}\left[
\({Q\over 2}-x\)^2 e^{-t}-{\sinh^2 \({Q\over 2}-x\){t\over 2} \over
\sinh{bt \over 2}\sinh {t\over 2b} }
\right]\right\}
}
and
\eqn{\Upso}{
\U_0\mathop{=}\limits^{\rm def}\left. {d\U(x)\over dx}\right|_{x=0}=\U(b)~.
}
An important property of the function $\U(x)$ is the set of recursion
relations it satisfies,
\eqn{\Uob}{\eqalign{
\U(x+b)&=\gamma(bx) b^{1-2bx}\U(x)~,
\cr
\U(x+1/b)&=\gamma(x/b) b^{2x/b-1}\U(x)~.
}}
Special cases of these relations that are used in the text are:
\eqn{\Ubb}{\eqalign{
{\U(2b)\over\U(b)}&=\gamma(b^2)b^{1-2b^2}~, \cr
{\U(2b)\over \U(2b-1/b)}&={\gamma(2-1/b^2)b^{3-2/b^2}}~, \cr
{\U(2b-R)\over\U(b-R)}&=\gamma(b(b-R))\, b^{1-2b^2+2bR}~,
\cr
{\U(2b-R)\over \U(2b-1/b-R)}&=\gamma(2-1/b^2-R/b)\, b^{3-2/b^2-2R/b}~.
}}
Using these relations one finds:
\eqn{\threeb}{\eqalign{
\langle V_{b}V_{b}V_{b}\rangle_{\rm sphere}&=
\left[\pi\mul \gamma(b^2)b^{2-2b^2}\right]^{1/b^2-2}{\U_0\over
\U(2b-1/b)}\({\U(2b)\over\U(b)}\)^3 \cr
&=
b^{-1}\left[\pi\mul \right]^{1/b^2-2}[\gamma(b^2)]^{1/b^2}\gamma(2-1/b^2)~,
}}
\eqn{\threel}{\eqalign{
\langle V_{b-{R\over 2}}V_{b-{R\over 2}}V_{b}\rangle_{\rm sphere}&=
\left[\pi\mul \gamma(b^2)b^{2-2b^2}\right]^{{1\over b^2}-2+{R\over b}}
{\U_0\U(2b)\U^2(2b-R)\over
\U\(2b-{1/ b}-R\)\U(b-R)\U^2(b)} \cr
&=b^{-1}\left[\pi\mul \gamma(b^2)\right]^{{1\over b^2}+{R\over b}-1}
\gamma(b^2) \gamma\(b(b-R)\) \gamma\(2-{1\over b^2}-{R\over b}\)~.
}}

\appendix{B}{ Solution of the equation for non-perturbative corrections
in the $c=1$ model.}

In this appendix we present the solution of equation {\eqg}.
First, differentiate equation {\eqg} once w.r.t. $y$. The result is
\eqn{\deqg}{
\left[ 2\sqrt{\alpha}\sinh\({2-R\over 2R }\XX\)\mp \cos ( \p_y g)\right]
\p^2_y g ={\sqrt{\alpha}\over 2}e^{\hf \XX}\p_y \XX(y\p_y-1)g=
\mp \hf\p_y \XX \sin(\p_y g)~.
}
Define
\eqn{\hg}{
h(\XX)=\pm\cos \(\p_y g(y)\)~.
}
In terms of $h$, equation \deqg\ reads
\eqn{\eqh}{
\left[ h - 2\sqrt{\alpha}\sinh\({2-R\over 2R }\XX\)\right] \p_{\XX}h
=\hf (h^2-1)~.
}
Changing variables to
\eqn\zzzzz{
z=\exp\( {2-R\over 2R}\XX\)
}
and considering $z$ as a function of $h$,
one arrives at a Riccati type equation
\eqn{\eqzh}{
\p_h z=a(h)z^2+b(h)z+c(h)~,
}
where
\eqn{\coefabc}{
a(h)=-{\sqrt{R-1}\over R}{1\over h^2-1}, \qquad
b(h)={2-R\over R}{h\over h^2-1}, \qquad
c(h)={\sqrt{R-1}\over R}{1\over h^2-1}~.
}
The general solution of \eqzh\ is easily written if a particular solution is known.
One can check that
\eqn{\solh}{
z_0(h)=-\sqrt{R-1}\(h+ \sqrt{h^2-1}\)
}
is a solution. Then after the substitution
\eqn{\zw}{
z(h)=z_0(h)+{1\over w(h)}~,
}
equation {\eqzh} takes the form
\eqn{\eqwh}{
\p_h w=-(2az_0+b)w-a~,
}
whose general solution is given by
\eqn{\solw}{
w(h)=-{1\over d(h)}\(C+\int dh\, a(h)d(h) \) \quad {\rm where}
\quad
d(h)=\exp \( \int dh\, (2az_0+b) \)~.
}
The integrals can be calculated explicitly. First, one finds
\eqn{\rh}{
d(h)=\sqrt{h^2-1}\,
\exp \left[ 2{R-1\over R}\log \(h+ \sqrt{h^2-1}\) \right]~.
}
The subsequent substitution into $w$ gives
\eqn{\solww}{
w(h)=-{1\over d(h)}\(C- {1\over 2\sqrt{R-1}}
\exp \left[2{R-1\over R}\log \(h+ \sqrt{h^2-1}\) \right] \)~.
}
Taking together {\zw}, {\solh} and {\solww}, one obtains
\eqn{\solzz}{
z(h)=-\sqrt{R-1}\left\{ h + \sqrt{h^2-1}\,
{C'\exp \left[-2{R-1\over R}\log \(h+ \sqrt{h^2-1}\) \right]+1
\over C'\exp \left[-2{R-1\over R}\log \(h+ \sqrt{h^2-1}\) \right]-1 }
 \right\}~,
}
where we redefined the integration constant $C'=2\sqrt{R-1} C$.

The solution {\solzz} is written in a form valid for $|h|>1$.
We are interested in $|h|<1$ due to
the definition {\hg}. Therefore, one should replace $\sqrt{h^2-1}\to
i\sqrt{1-h^2}$ in {\solzz}. The integration constant $C'$ must be chosen such that
$z(h)$ is real. This implies that $C'$ is a phase,  $C'=e^{2i\psi}$.
Denoting
\eqn\hhhh{
h=\pm \cos\phi, \quad (\phi=\p_y g(y))~,
}
one finds\foot{The ambiguity to replace $\phi\to \phi+2\pi n$
is captured by the integration constant $\psi$.}
\eqn{\solzphi}{
z(\phi)=\mp \sqrt{R-1}\left\{ \cos \phi +
\sin\phi \tan \( {R-1\over R} \phi+\psi\)
 \right\}~.
}
Finally, we rewrite {\solzphi} as
\eqn{\solzzz}{
z(\phi)=\mp\sqrt{R-1}\, { \cos \({1\over R}  \phi-\psi\)
\over \cos \( {R-1\over R} \phi+\psi\) }~.
}
The initial condition {\incon} means that $z\({\pi R}\)=0$,
and leads to
\eqn{\inpsi}{
\psi={\pi \over 2}~.
}
This reduces \solzzz\ to \zzz.
In terms of $z$ and $\phi$, our initial function $g$ is written as follows
\eqn{\gphi}{
g(\phi)=y(\phi)\phi\pm {1\over \sqrt{\alpha}}e^{-\hf \XX(\phi)}\sin \phi
=z^{-{R\over 2-R}}\left( (z^{-1}-z)\phi\pm {1\over \sqrt{\alpha}}\sin \phi
\right)~.
}
For the particular case $R=3/2$ the solution can be written explicitly.
Then \zzz\ and \wf\ give
\eqn\zzzr{\eqalign{
 e^{{1\over 6}X}&=\pm\sqrt{2}\cos{\phi\over 3}~,
\cr
e^{-1/3 X}&={1\over 2}(1+\sqrt{1+4y})
}}
and \gphi\ leads to
\eqn\gggr{
g(y)=3y\,\arccos\left[\pm \( 1+\sqrt{1+4y}\)^{-1/2}\right]
\pm {1\over 4}(1+4y)^{1/4}(3-\sqrt{1+4y})~.
}

\appendix{C}{Disk one-point functions in $c=1$ CFT}

We want to calculate the one-point function
$B_{\CT}=\langl\cos[R(x_L-x_R)]\rangl$ of the winding one
operator on the disk with Neumann boundary conditions for $x$. This
is the same as $\langl\cos( x/R)\rangl$ with Dirichlet boundary
conditions for $x$, which lives on a circle of radius $1/R$.

A simple way to calculate $B_{\CT}$ is to study
the annulus partition function with Neumann boundary
conditions, as a function of the modulus $\tau$ in the open string channel,
\eqn\ANNU{
Z_{ann}= \Tr q^{L_0-{1\over 24}}~,
}
where $q=e^{2\pi i \tau}$ and $L_0=p^2+N$. The momentum $p$ takes values
in ${1\over R}\IZ$. A standard calculation gives
\eqn\ANNUCA{  Z_{ann}={1\over \eta(q)}\sum_{n\in \IZ} q^{(n/R)^2}=
{\theta_3(0|{2\tau\over R^2})\over \eta(\tau)}~,
}
where $\eta(q)=q^{1\over 24}\prod_{m=1}^\infty (1-q^m)$ and
$\theta_3(0|\tau)=\sum_{n\in \IZ} q^{\hf n^2}$.

In the closed string channel the same annulus
partition function can be represented
as a sum over winding modes exchanged between two D-branes.
Performing the modular transformation $\tilde\tau=-1/\tau$, and using
the modular properties of the elliptic functions,
\eqn\MODUL{\eqalign{
\eta(\tau)&=(-i\tau)^{-1/2} \eta(-1/\tau)~, \cr
\theta_3(0|\tau)&= (-i\tau)^{-1/2} \theta_3(0|-1/\tau)~,
}}
we find the following expansion in the closed string channel
\eqn\CLOSEZ{
Z_{ann}={R\over \sqrt{2}} {1\over \eta(\tilde q)}
\sum_{n\in \IZ} \tilde q^{\hf(nR)^2}~,
}
where $\tilde q=\exp(-2\pi i/\tau)$. The disk partition sum \pfdisc\ is
the square root of the $n=0$ term  in \CLOSEZ. The one point function \bt\
is the square root of the $n=1$ term. Both are equal to
\eqn\BNOT{Z_{\rm Neumann}=B_{\CT}=2^{-{1\over4}}\sqrt{R}~.
}

\listrefs

\bye